\documentclass[preprint]{nature}
\usepackage{graphicx}
\usepackage{subfigure}
\usepackage{dcolumn}
\usepackage{bm}
\usepackage[mathlines]{lineno}
\usepackage{array}
\usepackage{braket}
\usepackage{diagbox}
\usepackage{appendix}
\usepackage{bm,color,bbm}

\usepackage{float}
\newcolumntype{.}{D{.}{.}{-1}}
\usepackage{amsopn}
\usepackage[colorlinks, linkcolor=red, anchorcolor=black, urlcolor=red, citecolor=blue]{hyperref}
\usepackage{epstopdf}

\usepackage{booktabs}
\usepackage{enumerate}
\usepackage{siunitx}
\usepackage{upgreek}
\usepackage{amssymb}
\usepackage{tabularx}
\usepackage{amsmath,amsthm,mathrsfs,amsfonts,dsfont}
\usepackage[10pt]{moresize}
\usepackage{amscd}
\usepackage{epsfig}
\usepackage{threeparttable}
\usepackage{multirow}
\usepackage[section]{placeins}
\usepackage{caption}

\newcommand{\tabincell}[2]{\begin{tabular}{@{}#1@{}}#2\end{tabular}}
\newcommand\mean[1]{\ensuremath{\langle #1 \rangle}}

\begin{document}
	\title{Microsatellite-based real-time quantum key distribution}

	\author
	{Yang Li$^{1,2,3}$, Wen-Qi Cai$^{1,2,3}$, Ji-Gang Ren$^{1,2,3}$, Chao-Ze Wang$^{1,2,3}$, Meng Yang$^{1,2,3}$, Liang Zhang$^{3,4}$, Hui-Ying Wu$^{5}$, Liang Chang$^{5}$, Jin-Cai Wu$^{3,4}$, Biao Jin$^{6}$, Hua-Jian Xue$^{1,2,3}$, Xue-Jiao Li$^{1,2,3}$, Hui Liu$^{6}$, Guang-Wen Yu$^{1,2,3}$, Xue-Ying Tao$^{1,2,3}$, Ting Chen$^{5}$, Chong-Fei Liu$^{3,4}$, Wen-Bin Luo$^{1,2,3}$, Jie Zhou$^{6}$, Hai-Lin Yong$^{6}$, Yu-Huai Li$^{1,2,3}$, Feng-Zhi Li$^{1,2,3}$, Cong Jiang$^{7}$, Hao-Ze Chen$^{8}$, Chao Wu$^{8}$, Xin-Hai Tong$^{9}$, Si-Jiang Xie$^{9}$, Fei Zhou$^{7}$, Wei-Yue Liu$^{1,2,3}$, Nai-Le Liu$^{1,2,3}$, Li Li$^{1,2,3}$, Feihu Xu$^{1,2,3}$, Yuan Cao$^{1,2,3}$, Juan Yin$^{1,2,3}$, Rong Shu$^{3,4}$, Xiang-Bin Wang$^{7}$, Qiang Zhang$^{1,2,3,7}$, Jian-Yu Wang$^{3,4}$, Sheng-Kai Liao$^{1,2,3}$, Cheng-Zhi Peng$^{1,2,3}$ and Jian-Wei Pan$^{1,2,3}$}
	\maketitle
		
	\begin{affiliations}
		\item Hefei National Research Center for Physical Sciences at the Microscale and School of Physical Sciences, University of Science and Technology of China, Hefei, China
		\item Shanghai Research Center for Quantum Science and CAS Center for Excellence in Quantum Information and Quantum Physics, University of Science and Technology of China, Shanghai, China
		\item Hefei National Laboratory, University of Science and Technology of China, Hefei, China
		\item Key Laboratory of Space Active Opto-Electronic Technology, Shanghai Institute of Technical Physics, Chinese Academy of Sciences, Shanghai, China
		\item Innovation Academy for Microsatellites of Chinese Academy of Sciences, Chinese Academy of Sciences, Shanghai, China
		\item Quantum CTek Co., Ltd., Hefei, China
		\item Jinan Institute of Quantum Technology, Jinan, China
		\item CAS Quantum Network Co., Ltd., Shanghai, China
		\item Beijing Electronics Science and Technology Institute, Beijing, China
	\end{affiliations}

	\begin{abstract}
A quantum network\cite{Kimble2008,Azuma2023} provides an infrastructure connecting quantum devices with revolutionary computing, sensing, and communication capabilities. As the best-known application of a quantum network, quantum key distribution (QKD) shares secure keys guaranteed by the laws of quantum mechanics. A quantum satellite constellation offers a solution to facilitate the quantum network on a global scale\cite{Lu2022RMP,2017satelliteProgress}. The Micius satellite has verified the feasibility of satellite quantum communications\cite{Liao2017Nature,Yin2017Science,ren2017ground,Liao2018PRL}, however, scaling up quantum satellite constellations is challenging, requiring small lightweight satellites, portable ground stations and real-time secure key exchange.
Here we tackle these challenges and report the development of a quantum microsatellite capable of performing space-to-ground QKD using portable ground stations. The quantum microsatellite features a payload weighing approximately 23 kg, while the portable ground station weighs about 100 kg. These weights represent reductions by more than an order and two orders of magnitude, respectively, compared to the Micius satellite\cite{Liao2017Nature,Yin2017Science,ren2017ground,Liao2018PRL}. Additionally, we multiplex bidirectional satellite-ground optical communication with quantum communication, enabling key distillation and secure communication in real time.
Using the microsatellite and the portable ground stations, we demonstrate satellite-based QKD\cite{Bennett1984,ekert1991quantum} with multiple ground stations and achieve the sharing of up to 0.59 million bits of secure keys during a single satellite pass. The compact quantum payload can be readily assembled on existing space stations\cite{NASA2,gu2022china} or small satellites\cite{albulet2016spacex}, paving the way for a satellite-constellation-based quantum and classical network for widespread real-life applications.
	\end{abstract}
		
	\maketitle

\section*{Introduction}
The quantum internet\cite{Kimble2008,Azuma2023} promises to interlink quantum devices at distant nodes, harnessing the transformative capabilities of quantum information processing.
As the quintessential application of a quantum internet, quantum key distribution (QKD)\cite{Bennett1984,ekert1991quantum} offers a secure means of communication based on the fundamental laws of physics\cite{xu2019quantum}. Substantial progress has been made in extending the transmission distance through optical fibers\cite{Peng2007RPL,rosenberg2007longPRL,Liu2023PRL} or terrestrial free-space\cite{Bennett1989ACS, peng2005, Manderbach2007PRL, Ursin2007NP}. Particularly, satellite-based QKD provides a promising solution for global quantum communication. By using the Micius satellite, the satellite-ground quantum communications\cite{Liao2017Nature,Liao2018PRL,Yin2017Science,ren2017ground,Chen2021Nature} have been extensively verified, representing an essential step toward a global-scale quantum network\cite{2017satelliteProgress,Lu2022RMP}.

Till now, the demonstration of satellite-based QKD has employed large-size scientific satellites and massive optical ground stations (OGSs)\cite{Liao2017Nature, Liao2017CPL}, which poses challenges for constructing a practical quantum satellite constellation. This motivates the development of lightweight and low-cost quantum satellites. Indeed, several proposals\cite{Oi2017EQT,Neumann2018EQT,Kerstel2018EQT,Haber2018AACSS,Podmore2019,Miller2023,Ahmadi2024,Mazzarella2020} and preliminary test\cite{Villar2020Optica} for CubeSats and quantum constellations have been reported in recent years. Despite the progress, the CubeSat proposal remains to be verified in experiments, especially with small-aperture and portable OGSs to facilitate general applications. On the other hand, the extraction of secure keys necessitates a series of classical communication for key distillation. This was predominantly accomplished through conventional microwave communication\cite{Liao2017Nature,Liao2018PRL}. However, the limited bandwidth and communication time have resulted in long time delays in distilling the secure keys.

Here we tackle these challenges and report the developments of a quantum microsatellite, a portable ground station, and multiplexed satellite-ground quantum and optical communications for real-time key extraction. We have built a compact 625-MHz QKD light source by leveraging a single laser diode (LD), enhancing both the integration and practical security\cite{Liao2017Nature}. Additionally, we have implemented a $\sim\mu$rad-precision tracking technology integrated with satellite attitude control, leading to a substantial reduction in the weight of the satellite's acquisition, pointing and tracking (APT) system. With these achievements, we have launched the quantum microsatellite --- Jinan-1 --- into a 500-km Sun-synchronous orbit on July 27, 2022. The payload merely weighs 23 kg, which represent a reduction exceeding an order of magnitude compared to the payload mass of the Micius satellite\cite{Liao2017Nature,Yin2017Science,ren2017ground,Liao2018PRL}. The OGSs have also been minimized with weight from an approximate 13,000 kg to a mere 100 kg, offering the advantage of flexible and rapid deployment in complicated urban and mountainous environments. Meanwhile, we implement bidirectional optical communication between the satellite and OGSs, enabling real-time secure key distillation and secure communication. Using Jinan-1 microsatellite, satellite-to-ground QKD experiments (see Fig.~\ref{fig_setup}a) have been effectively demonstrated with multiple portable OGSs in the urban cities of Jinan (36$^{\circ}$40'38''N, 117$^{\circ}$7'21''E), Hefei (31$^{\circ}$49'37''N, 117$^{\circ}$8'27''E), Wuhan (30$^{\circ}$28'35''N, 114$^{\circ}$31'26''E) and in the mountainous area of Nanshan (43$^{\circ}$28'31''N, 87$^{\circ}$10'36''E). As an example, during a satellite pass on 25 September 2022, a total of 406784-bit keys were shared in real time between the satellite and the OGS.

\section*{Compact QKD light source with a single laser diode}
\begin{figure}[htbp]
\centering
\includegraphics[width =0.8 \linewidth]{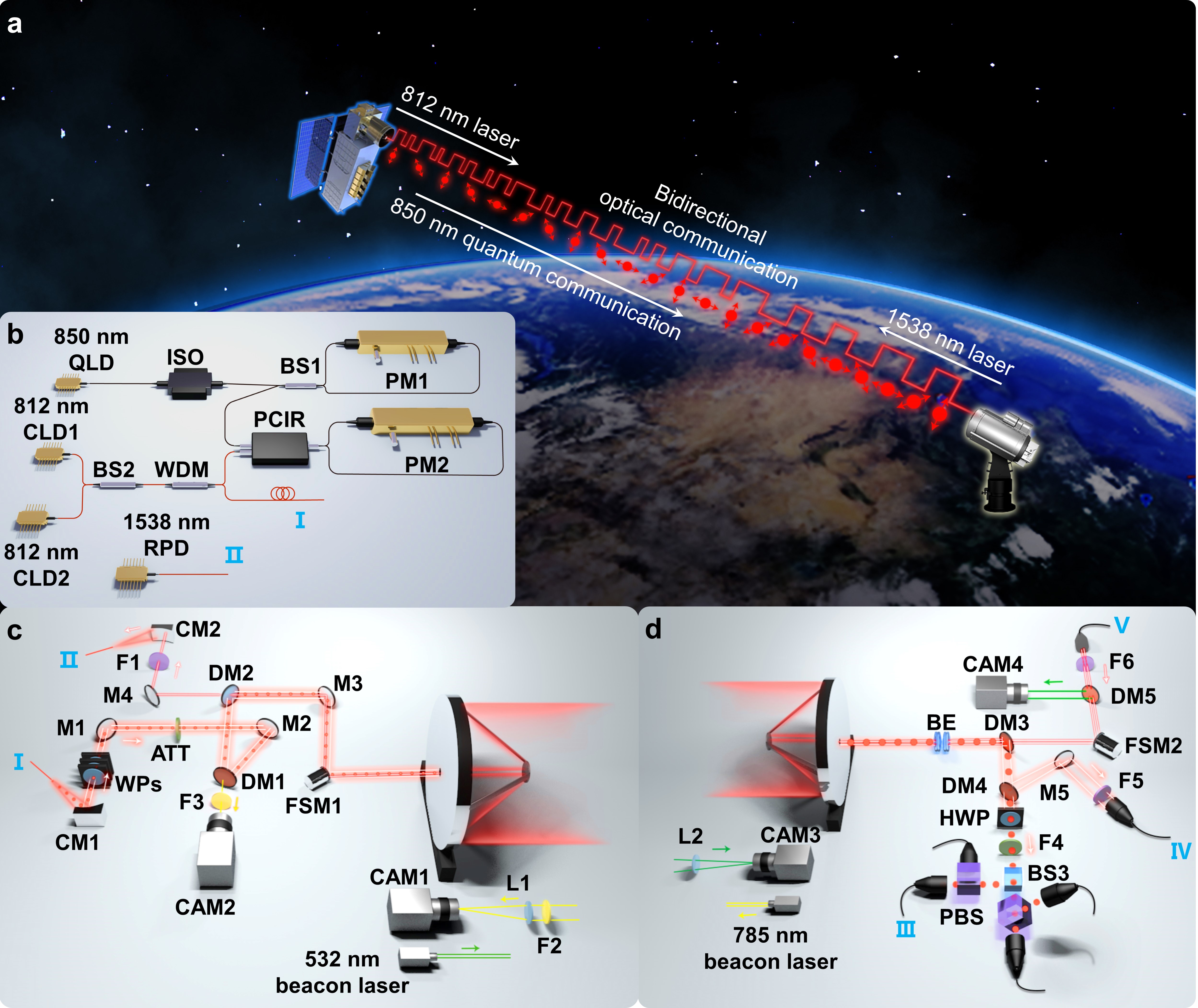}
\captionsetup{font={footnotesize}}
\caption{ 
\textbf{The experimental setup.}
\textbf{a}, Overview of the satellite-to-ground quantum key distribution (QKD) experiment. In addition to the downlink 850 nm quantum photons, the satellite and the optical ground station (OGS) are equipped with bidirectional optical communication (812 nm downlink laser and 1538 nm uplink laser), enabling key distillation and secure communication in real time.
\textbf{b}, QKD light source. The QKD light source uses a single laser diode (LD) and external modulations. QLD, quantum LD; CLD, communication LD; ISO, isolator; BS, beam splitter; PM, phase modulator; PCIR, polarization module; RPD, receiving photodiode; WDM, wavelength division multiplexer.
\textbf{c}, Satellite optical design. The labels \uppercase\expandafter{\romannumeral1} and \uppercase\expandafter{\romannumeral2} represent the output single-mode fiber (SMF) of the 812 nm laser and 850 nm quantum light, and the receiving multimode fiber (MMF) for 1538 nm laser, respectively. CM, concave mirror; WP, wave plate; M, mirror; ATT, attenuator; DM, dichromic mirror; CAM, camera; F, filter; L, lens; FSM, fast steering mirror.
\textbf{d}, Portable OGS.
The labels \uppercase\expandafter{\romannumeral3}, \uppercase\expandafter{\romannumeral4}, and \uppercase\expandafter{\romannumeral5} represent the receiving MMFs for 850 nm quantum photons, receiving MMF for 812 nm laser, and transmitting SMF for 1538 nm laser, respectively. BE, beam expander; PBS, polarization beam splitter; HWP, half-wave plate.
}
\label{fig_setup}
\end{figure}

We utilize a single LD with external modulations to implement the decoy-state QKD light source (see Fig.~\ref{fig_setup}b). An 850 nm LD is gain-switched to emit consecutive short optical pulses at a repetition frequency of 625 MHz. After an isolator, the intensity modulation module consisting of a beam splitter (BS) and a lithium niobate ($LiNbO_{3}$) phase modulator (PM) is used to prepare the intensity states (signal state, decoy state, and vacuum state) of the decoy-state theory\cite{Wang2005PRL, Lo2005PRL} using a Sagnac interferometer scheme\cite{Roberts2018OL}. Similarly, a customized polarization module and another PM are employed to prepare the polarization states of BB84 protocol\cite{Bennett1984}, also based on a Sagnac interferometer scheme\cite{Li2019OL}. In comparison to the method employing multiple LDs\cite{Liao2017Nature}, this Sagnac-interferometer-based modulation scheme enables a much higher repetition rate while ensuring inherent robustness for complex spaceborne applications. The ground test results demonstrated an extinction ratio of approximately 29 dB and an average polarization contrast ratio of approximately 25 dB. By utilizing a single LD and outputting through a single-mode fiber (SMF), the inherent uniformity across other photon dimensions, such as space, time, and frequency can be ensured, effectively mitigating the risk of potential side-channel information leakage. The integrated design of optical components, drive electronics, and the fusion of optical fiber components contribute to the high integration and light weight of the QKD light source.

Utilizing an SMF to guide the quantum photons from the QKD light source to the transmitting telescope introduces implications for the prepared polarization states, necessitating compensation measures. We employ a combination of two motorized quarter-wave plates (QWPs) and one motorized half-wave plate (HWP) (see Fig. \ref{fig_setup}c) to compensate for the corresponding polarization transformation (see Methods), and another motorized HWP at the OGS (see Fig.~\ref{fig_setup}d) to compensate for the time-dependent polarization rotation caused by the satellite-OGS relative motion\cite{Liao2017Nature}. The transmitted vacuum state can be utilized to assess the impact of induced additional quantum bit error rate (QBER) due to dark counts. The normalized QBER, factoring in elements such as the QKD light source, satellite and ground telescopes, and polarization compensations, fluctuates between zero and 1.0\% during the whole orbit (see Fig. \ref{fig_Result}a). This result closely resembles the polarization preparation fidelity of the QKD light source, verifying the resilience of the polarization compensation method.

\section*{High-precision tracking based on satellite attitude control}
\begin{figure}[htbp]
\centering
\includegraphics[width =1 \linewidth]{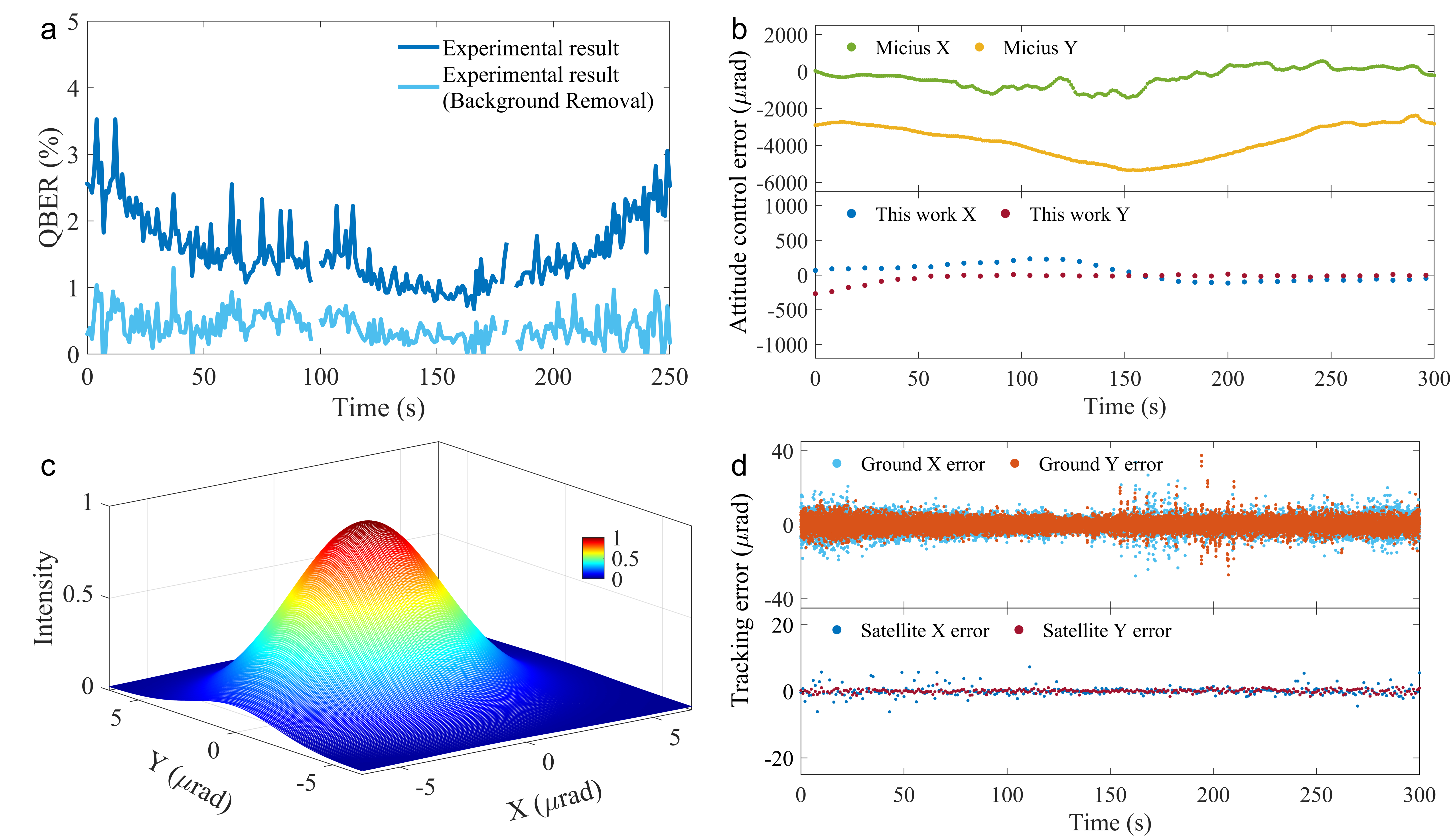}
\captionsetup{font={footnotesize}}
\caption{
\textbf{Characterization of the satellite-to-ground QKD system.}
\textbf{a}, Quantum bit error rate (QBER) performance. After removing the induced additional QBER due to background noise, the normalized QBER exhibits a strong concordance with the polarization preparation fidelity of the QKD light source (approximately 25 dB).
\textbf{b}, Distribution of the attitude control error of the Micius satellite\cite{Liao2017Nature} and this microsatellite. The measured attitude control error range of this microsatellite is approximately 280$\sim$350 $\mu rad$, showcasing a substantial improvement compared to the 2.0$\sim$3.0 $mrad$ of the Micius satellite and sufficient to meet the requirement for fine tracking. Owing to data bandwidth constraints, the update rate of attitude telemetry in this microsatellite is 1/8 Hz, resulting in data updates occurring every 8 seconds.
\textbf{c}, Measured far-field pattern from the satellite-to-ground scanning test. The measured divergence angle (full angle at $1/e^{2}$ maximum) is 9 $\sim10~{\mu}rad$, rendering QKD applications viable with small-aperture portable OGSs.
\textbf{d}, Tracking performances of the satellite and the portable OGS. The final achieved tracking precisions (root mean square) are approximately $0.55\sim1.6~\mu$rad for the satellite, and approximately $3.3\sim4.0~\mu$rad for the OGS, adequately fulfilling the requirements for quantum communication and classical communication.
}
\label{fig_Result}
\end{figure}

Satellite-to-ground QKD demands stringent link efficiencies, which is quadratically proportional to the apertures of the telescopes used by the satellite and the ground station. Given the practical necessity for large-amount ground users in real-world scenarios, the meter-size telescopes employed in the Micius experiment\cite{Liao2017Nature,Yin2017Science,ren2017ground,Liao2018PRL} are inadequate for these requirements.
Here we develop a satellite-borne 200-mm telescope, which is used in conjunction with the ground station's 280-mm aperture telescope to achieve satellite-ground quantum communication. 
This choice enables a near-diffraction-limited divergence angle of approximately 9 $\sim10~{\mu}rad$ (see Fig. \ref{fig_Result}c) for the quantum photons, and ensures the miniaturization and portability of the ground station.
In contrast, previous proposals of CubeSats\cite{Oi2017EQT,Neumann2018EQT,Kerstel2018EQT,Haber2018AACSS,Podmore2019,Miller2023,Ahmadi2024} generally incorporate telescopes of smaller diameters, which requires ground telescopes with much larger apertures, limiting their real-life applications.

When using a 200-mm-aperture telescope, the miniaturization of the APT system poses challenges due to its bulky opto-mechanical structure. Previous demonstrations of high-precision APT systems typically employ a two-stage scheme\cite{Liao2017Nature, Liao2017CPL} involving a complex two-axis mirror or mount for coarse tracking. In our microsatellite, however, we introduce satellite attitude control for coarse tracking and simplify the two-axis mirror (or mount) and the associated rotating mechanism, effectively reducing the weight of the APT system. To improve the precision of satellite attitude control, we incorporate the detected uplink beacon laser of the capture camera into the closed loop of satellite attitude control and increase the frequency of attitude control to 20 Hz. Considering atmospheric turbulence, clouds, and background light, the capture camera's output may occasionally be compromised or present false targets. To address this, we employ a dual-source arbitration strategy, which can flexibly switch between star-sensor-based and capture-camera-based attitude control modes, ensuring the continuous and robust control of the satellite's attitude (see Methods).

The final measured attitude control error range is approximately 280$\sim$350 $\mu rad$ (see Fig. \ref{fig_Result}b), sufficient to direct the uplink beacon light into the FOV of the fine camera. This result is approximately one order of magnitude lower than the 2.0$\sim$3.0 $mrad$  attitude control error range of the Micius satellite. Building upon this achievement, with feedback control from the high-precision fine camera and the high-bandwidth piezoelectric fast steering mirror (FSM), a final tracking precision (root mean square) of $0.55\sim1.6~\mu$rad was achieved (see Fig. \ref{fig_Result}d), meeting the demands for both quantum light transmission and classical communication of the microsatellite.

\section*{Implementation of the microsatellite}
While achieving a compact 625-MHz QKD light source, $9\sim10~{\mu}rad$ far-field divergence angle and $\sim\mu$rad tracking precision, the weight of the payloads has been significantly reduced to 22.7 kg. In comparison to the ${\sim}$250 kg of Micius satellite, the payload weight has been reduced by over an order of magnitude. Based on the payloads, we adopt a modular approach to realize the satellite with an overall weight of 95.9 kg (see Fig. \ref{fig_satellite}) (see Methods). This microsatellite and the lightweight payloads not only enable quantum satellite constellation networking through batch launches, but also opens up integration possibilities with space station\cite{NASA2,gu2022china} and existing satellite internet constellations\cite{albulet2016spacex}.

\section*{Portable optical ground station}
\begin{figure}[htbp]
\centering
\includegraphics[width =0.8 \linewidth]{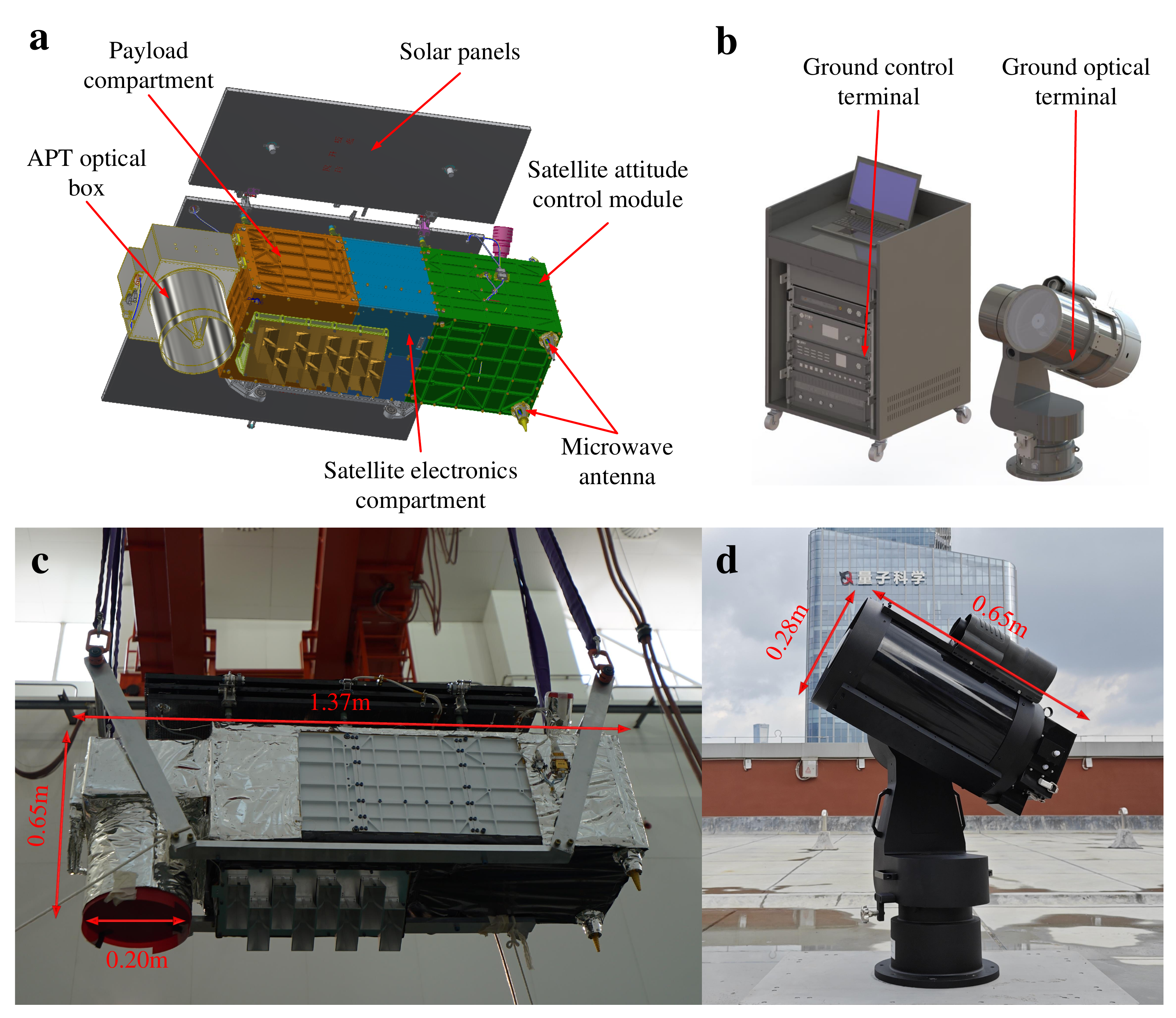}
\caption{
\textbf{The microsatellite and the portable OGS}.
\textbf{a}, The satellite consists of an APT optical box, payload compartment, satellite electronics compartment, satellite attitude control module, microwave antenna and solar panels.
\textbf{b}, The OGS consists of a control terminal and an optical terminal.
\textbf{c}, Photo of the microsatellite before being assembled into the rocket. The satellite launch state envelope measures approximately 1.37 m $\times$ 0.49 m $\times $0.65 m, with a telescope aperture of 0.2 m. 
\textbf{d}, Photo of the portable OGS in the urban area of Jinan. The main telescope of the OGS has an envelope size of approximately 0.65 m $\times$ 0.28 m $\times $0.28 m.
}
\label{fig_satellite}
\end{figure}

The portable OGS (see Fig. \ref{fig_satellite}) is composed of a control terminal and an optical terminal. The control terminal houses electronics for the telescope and gimbal mount control, along with the ground QKD device.
The optical terminal comprises a two-axis gimbal mount, a main telescope, followed by a beam expander, dichromic mirrors, an HWP for dynamical polarization compensation, 4-nm band-pass filters for suppressing background noise, and the BB84 module for polarization state decoding.
The overall efficiency of the OGS, as determined by indoor testing, is approximately 49\%. The received quantum photons are coupled into four 105-$\mu$m multimode fibers (MMFs) and detected by four single-photon detectors (SPDs) with efficiencies of $\sim$ 60\% and full-width-at-half-maximum (FWHM) timing jitter less than 350 ps. Through an optical-mechanical integrated design, the weight of the optical terminal has been minimized from approximately 13000 kg to a mere 100 kg, representing a reduction exceeding two order of magnitude, and the installation time of the OGS has been streamlined to approximately $3\sim5$ hours. This makes it highly adaptable for flexible applications and swift deployment in diverse complex scenarios.

\section*{Multiplexed quantum and classical communication}
To enhance the timeliness of key distillation, we have integrated bidirectional optical communication between the satellite and the OGSs. The downlink communication laser operates at a wavelength of $\sim$812 nm, while the uplink communication laser operates at $\sim$1538 nm, with a communication code rate of 156 Mbps. The 812 nm laser and the 850 nm quantum light are combined using a fiber wavelength division multiplexer and then output through the same SMF, ensuring efficient optical alignment and coaxiality. A customized attenuator is employed to attenuate the 850 nm quantum light to the single-photon level while maintaining a high transmittance of the 812 nm downlink laser. By realizing an output power of approximately 2.4 mW and a narrow divergence angle of $9 \sim10~{\mu}rad$, we eliminate the use of high-power optical amplifier for the downlink 812 nm laser, reducing both power consumption and weight requirement. The uplink communication laser is detected by a linear-mode avalanche photodiode with a response sensitivity of approximately -35 dBm.

To achieve high-precision time synchronization, specific synchronization words with a repetition frequency of $\sim$9.5 kHz are encoded into the downlink communication laser. At the ground receiver, the clock and communication data are recovered using the clock data recovery technique. Once the data are decoded, the target synchronization words can be extracted for time synchronization between the satellite and the OGS. Ground test results demonstrated a synchronization precision (FWHM) of approximately 100 ps. Taking into account the optical pulse width, time synchronization precision, SPD jitter, and time-to-digital converter precision, a temporal gating width of 800 ps is employed, which can suppress the background noise by 3 dB and ensure a $\sim$80\% gating efficiency for quantum photons.

\section*{Experimental results}
The satellite follows a Sun-synchronous orbit, completing a pass over the OGS once every night, commencing at approximately 22:30 local time. Since September 2022, we have successfully conducted satellite-to-ground QKD experiments under favourable atmospheric conditions. In this process, the satellite's QKD source utilizes physical random numbers generated by a chip\cite{qian2017note} to prepare signal state, decoy state, and vacuum state with respective ratios of 0.5, 0.25, and 0.25 and average photon numbers of approximately 0.8, 0.1, and 0.001. The key distillation process employs the decoy-state QKD analysis method\cite{Wang2005PRL, Lo2005PRL} with the failure probability set to $10^{-10}$ using a Chernoff bound to account for statistical fluctuations\cite{chernoff1952measure, jiang2017measurement} (see Methods). Through the utilization of bidirectional optical communication for time synchronization and data post-processing, the continuous real-time distribution of quantum photons and distillation of secure keys are achieved.

\begin{figure}[htbp]
\centering
\includegraphics[width =0.7 \linewidth]{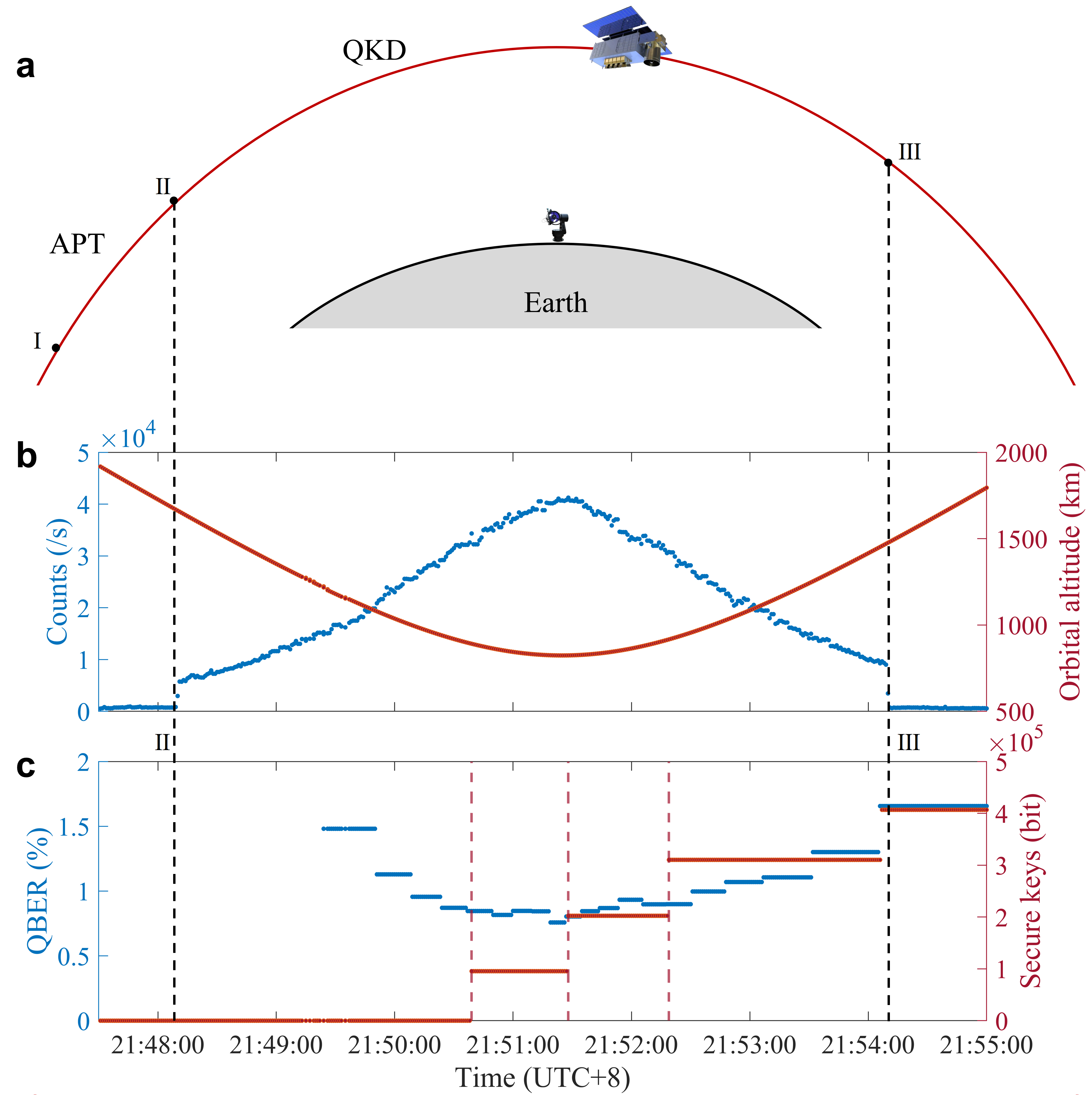}
\captionsetup{font={small}}
\caption{ 
\textbf{Experimental procedure of satellite-to-ground QKD, and the experimental results during a single orbit on 25 September 2022.}
\textbf{a}, Experimental procedure of a satellite orbit. Approximately 5 minutes before entering the shadow zone, the satellite utilizes flywheels to control its attitude and starts APT (label \uppercase\expandafter{\romannumeral1}). When the satellite reaches an elevation angle of $10{\sim}15^{\circ}$ from the horizon, the optical link is firmly established, marking the beginning of QKD (label \uppercase\expandafter{\romannumeral1}). The experimental duration for each satellite orbit is approximately 6 minutes, and the label \uppercase\expandafter{\romannumeral3} represents the end of QKD.
\textbf{b}, The variation in physical distance from the satellite to the station and the corresponding detected photon count rate over time.
\textbf{c}, The measured QBER of the sifted key in each 100-kbit packet and the secure keys extracted from every five packets as a function of time. During the QKD procedure, once the required number of quantum photon detection instances has been accumulated, the final secure key can be generated in real-time.
}
\label{fig_QKD}
\end{figure}

For one satellite pass on 25 September 2022, we placed the ground station on the rooftop of a building in downtown Jinan and conducted QKD experiments in real-life bright environments. 
The variation in physical distance between the satellite and the station, as well as the corresponding detected photon counts per second, are presented in Fig. \ref{fig_QKD}b.
During the data collection period lasting 300 seconds, a total of 20 packets of sifted keys were obtained, with each packet containing 100 kbits. The QBER for each packet ranged from 0.76\% to 1.79\% (see Fig. \ref{fig_QKD}c). In addition to system polarization performance, the QBER is more susceptible to dark counts originating from ambient ground-scattered light and detector noise. In the key distillation process, the final key is extracted in privacy-amplification packages of 500-kbits, which represents a compromise between single-orbit detection events and key distillation efficiency. After key distillation, a total of 406784 bits of secure keys were successfully shared between the satellite and Jinan station. Additional experimental results of satellite-to-ground QKD are provided in Table \ref{tab1_key}, revealing that the maximum number of secure keys obtained per orbit reached 592384 bits on 31 August 2023. After establishing QKD between the satellite and an OGS, the satellite can serve as a "space postman", enabling key relay and encrypted communication with a remarkably low latency of approximately 1.5 hours between two distant OGSs (see Methods).

\section*{Conclusion}
In summary, we have achieved remarkable success in designing, developing, and launching a lightweight microsatellite with a payload weight of 22.7-kg, significantly lighter than conventional satellites. Utilizing this microsatellite, we have successfully demonstrated real-time satellite-to-ground QKD with 100-kg class portable OGSs located in the urban areas of Jinan, Hefei, Wuhan and Nanshan. This achievement serves as a strong foundation for our future plans of launching multiple microsatellites and establishing a vast network of OGSs, leading us toward the realization of a practical quantum constellation that can offer quantum access services to users worldwide. The results not only lay the foundation for practically useful and large-scale quantum communications but also hold immense promise for the widespread deployment of quantum networks.

Further efforts can be devoted to integrated QKD light sources based on photonics chips\cite{Paraiso2021} to minimize the payloads, daytime satellite-to-ground QKD\cite{Liao2017NP} to provide services covering the entire 24 hours of a day, combining satellites at different altitudes\cite{Lu2022RMP} and diverse orbit types\cite{Li2022Optica} to achieve optimal complementarity, and exploring the feasibility of satellite-based measurement-device-independent QKD\cite{Lo2012PRL} and twin-field QKD\cite{Lucamarini2018Nature}.
Furthermore, with the development of quantum nodes and quantum repeaters, a global-scale quantum network\cite{Kimble2008,Azuma2023} with revolutionary communication\cite{duan2001long}, computing\cite{Ladd2010} and sensing\cite{degen2017quantum} capabilities are anticipated in the near future.

	\paragraph{Data availability}
	The datasets generated in the current study are available from the corresponding author upon reasonable request.
	
	\paragraph{Acknowledgments}
The authors thank C.-W. Feng, W.-S. Tang, J.-S. Dai, T.-T. Wang, T. Tang, Y.-C. Ji, S.-Q. Fan and Z. Wang for their efforts in the development of the payloads and the microsatellite, thank H.-B. Li, Z. Wang, W.-W. Ye, S.-J. Xu, X. Li, X. Han, Z.-G. Xiao, L.-K. Guo, C.-F Zhu and Y.-Y. Wang for their long-term assistance in observations and experimental measurements, thank Q.-Y. Yao, S.-Q. Zhao  and Y.-G. Zhao, for their efforts in the development of the portable OGSs, and thank Z.-Y. Chen and Y. Zheng for their help in drawing pictures. This work was supported by Jinan Innovation Zone, and Key R \& D Plan of Shandong Province (2021ZDPT01), National Key Research and Development Program of China (2020YFA0309701, 2020YFA0309703, 2022YFF0610100), Shanghai Municipal Science and Technology Major Project (2019SHZDZX01), National Natural Science Foundation of China (12174374, 12274398, 12374475) and Innovation Program for Quantum Science and Technology (2021ZD0300104).

\paragraph{References}
\bibliographystyle{naturemag_noURL}
\bibliography{Microsatellite_QKD}
\newpage
\clearpage

\clearpage
\clearpage

\clearpage
\clearpage

\clearpage
\clearpage

\clearpage
\clearpage

\clearpage
\begin{table}
\footnotesize
\centering
\begin{threeparttable}[b]
\caption{
\textbf{Experimental results of satellite-to-ground QKD from 20 September 2022 to 25 July 2023 based on Coordinated Universal Time.}
$R_{sifted}$ is the packet of sifted keys, $E_{\mu}$ is the sampling QBER of the signal states, and $R_{final}$ is the number of final secure keys.
The $E_{\mu}$ gives the range and the average of measurement results in packets. Ave, average; Max, maximum; Min, minimum.
}
\label{tab1_key}
\begin{tabular}{ccccccc}
\hline
Date& \tabincell{c}{Ground station \\ \& type} &\tabincell{c}{Maximum \\Elevation ($^{\circ}$)}& \tabincell{c}{$R_{sifted}$ \\(*100 kbits)} & Ave (Min, Max) of $E_{\mu}$ & \tabincell{c}{$R_{final}$ \\(bits)}\\
\hline
20 Sep. 2022 &Jinan \& Urban & 41.8  & 17 & 0.99\% (0.68\%, 1.60\%) & 330240\\
\hline
25 Sep. 2022 &Jinan \& Urban & 34.1  & 20 & 1.01\% (0.76\%, 1.79\%) & 406784\\
\hline
28 Sep. 2022 &Jinan \& Urban & 77.7  & 31 & 0.99\% (0.72\%, 2.03\%) & 571392\\
\hline
29 Sep. 2022 &Jinan \& Urban & 52.7  & 20 & 0.86\% (0.53\%, 1.30\%) & 436096\\
\hline
20 Feb. 2023 &Hefei \& Urban & 89.3  & 18 & 1.37\% (0.83\%, 2.92\%) & 264448 \\
\hline
4 Mar. 2023 &Hefei \& Urban & 76.3  & 18 & 1.48\% (0.93\%, 3.29\%) & 275328 \\
\hline
13 Mar. 2023 &Hefei \& Urban & 46.2 & 16 & 1.34\% (0.91\%, 2.66\%) & 262656 \\
\hline
1 Jul. 2023 &Nanshan \&  Mountainous & 84.7  & 26 & 1.66\% (0.96\%, 3.23\%) & 392320 \\
\hline
11 Jul. 2023 &Nanshan \&  Mountainous & 63.9  & 27 & 1.17\% (0.89\%, 1.73\%) & 538368 \\
\hline
12 Jul. 2023 &Nanshan \&  Mountainous & 54.5  & 28 & 1.58\% (0.86\%, 3.02\%) & 397952 \\
\hline
15 Jul. 2023 &Nanshan \&  Mountainous & 81.9 & 23 & 1.33\% (0.76\%, 2.76\%) & 382720 \\
\hline
18 Jul. 2023 &Nanshan \&  Mountainous & 67.2 & 27 & 1.00\% (0.75\%, 1.47\%) & 578048 \\
\hline
19 Jul. 2023 &Nanshan \&  Mountainous & 51.1 & 16 & 1.16\% (0.62\%, 1.70\%) & 337152 \\
\hline
22 Jul. 2023 &Nanshan \&  Mountainous & 75.9 & 32 & 1.41\% (0.69\%, 2.74\%) & 519552 \\
\hline
25 Jul. 2023 &Nanshan \&  Mountainous & 73.8 & 20 & 0.92\% (0.65\%, 1.45\%) & 447744 \\
\hline
26 Jul. 2023 &Nanshan \&  Mountainous & 46.1 & 24 & 1.23\% (0.65\%, 1.83\%) & 398208 \\
\hline
31 Aug. 2023 &Wuhan \&  Urban & 60.9 & 37 & 1.55\% (1.04\%, 3.93\%) & 592384 \\
\hline
1 Sep. 2023 &Wuhan \&  Urban & 45.2 & 33 & 2.33\% (1.40\%, 9.69\%)\tnote{1} & 321536 \\
\hline
\end{tabular}
\begin{tablenotes}
\item[1] During the experiment, the moon entered the field of view of the ground telescope, causing a significant increase in dark counts and consequently raising the QBER of the sifted key packet.
\end{tablenotes}
\end{threeparttable}
\end{table}
\clearpage

\section*{Methods}
\setcounter{figure}{0}
\setcounter{table}{0}
\captionsetup[figure]{name=Extended Data Figure}
\captionsetup[table]{name=Extended Data Table}

\subsection{Polarization compensation of quantum photons.}
To address the issues of polarization transformation of the SMF guiding the quantum photons from the QKD light source to the transmitting telescope, and the time-dependent rotation of photon polarization caused by the relative motion between the satellite and the OGS, we have implemented a two-step compensation approach. 
Firstly, we calculate the rotation angle offset by taking into account the relative motion of the satellite and the OGS and all the birefringent elements in the optical path (except the transmitting SMF).
A motorized HWP is then employed at the OGS to compensate for the polarization basis rotation.
This results in a stable polarization transformation between the QKD source at the satellite and the polarization decoding module at the OGS.
Under the condition that the polarization basis rotation has been compensated, we can precisely measure and compensate for the polarization change caused by the SMF.

\begin{figure}[htbp]
 \centering
 \includegraphics[width =0.9 \linewidth]{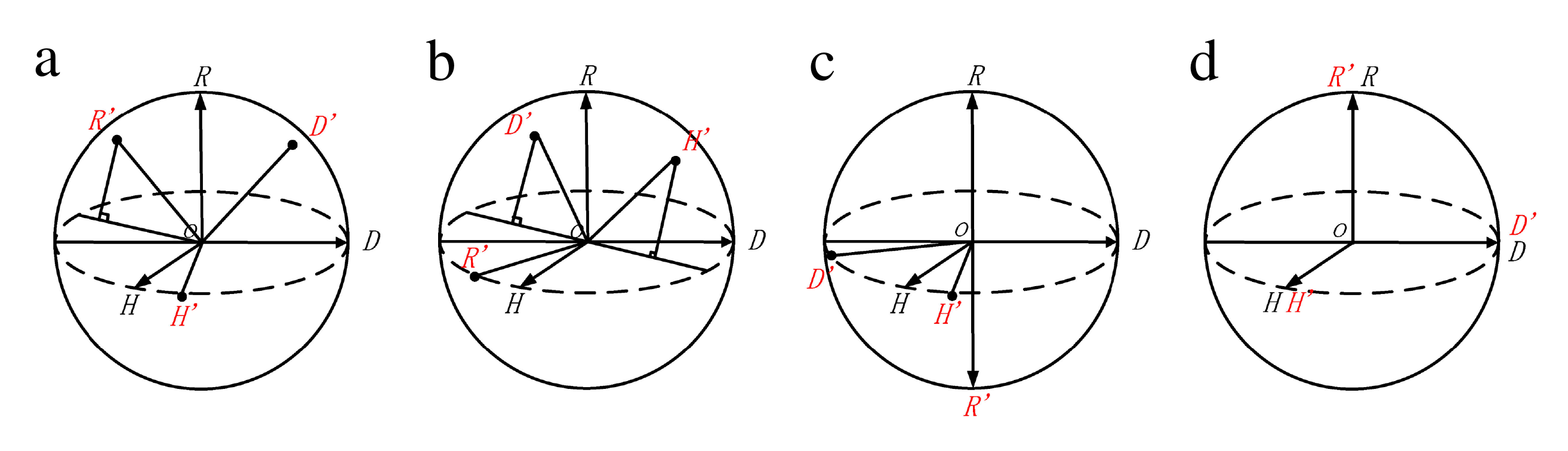}
 \caption{
\textbf{Schematic of the polarization compensation process on the Poincar\'{e} sphere}.
\textbf{a}, The first QWP moves the corresponding point $R^{'}$ to the equatorial plane, where the corresponding points $D^{'}$ and $H^{'}$ will appear on a warp coil. The central axis for rotation is the projection axis of $oR^{'}$ ($o$ is the original point) on the equatorial plane.
\textbf{b}, The second QWP moves the corresponding point $H^{'}$ to the equatorial plane, where the corresponding point $R^{'}$ will be in the pole position. The central axis for rotation is the projection axis of $oH^{'}$ on the equatorial plane.
\textbf{c}, The HWP moves the corresponding point $H^{'}$ to the original position of point $H$. The central axis for rotation is the middle axis of $oH^{'}$ and $oH$.
\textbf{d}, The corresponding points $R^{'}$, $H^{'}$ and $D^{'}$ will return to their original positions of $R$, $H$ and $D$ ($R^{'}$ coincides with $R$, $H^{'}$ coincides with $H$, $D^{'}$ coincides with $D$. The three wave plates help realize the previous unitary transformation process $U^{'}$. 
 }
\label{fig_polarization}
\end{figure}

By transmitting distinct quantum states ($\left|D\right\rangle=\frac{1}{\sqrt{2}}(\left|H\right\rangle+\left|V\right\rangle)$,
$\left|A\right\rangle=\frac{1}{\sqrt{2}}(\left|H\right\rangle-\left|V\right\rangle)$,
$\left|L\right\rangle = \frac{1}{\sqrt{2}} (\left|H\right\rangle +i \left|V\right\rangle)$,
and $\left|R\right\rangle = \frac{1}{\sqrt{2}} (\left|H\right\rangle - i\left|V\right\rangle)$) from the QKD light source, employing the set of three motorized wave plates (two QWPs and one HWP) at the satellite for basis alternatiton, and utilizing the BB84 module for polarization states decoding, we can deduce the transformation matrix of the SMF and the corresponding compensation matrix.
Noting that the relative positions of the corresponding points $R^{'}$, $H^{'}$ and $D^{'}$ (the new positions corresponding to the original $R$, $H$, and $D$ after the SMF) remain unchanged on the Poincar\'{e} sphere.
The set of three motorized wave plates is also adopted to compensate for the polarization transformation of the SMF, which can transform the $R^{'}$, $H^{'}$ and $D^{'}$ back to the original points $R$, $H$, and $D$ on the Poincar\'{e} sphere (see Extended Data Fig. \ref{fig_polarization}).

\subsection{Parameter and performance of the APT systems.}
The optical transmitter in the satellite and the receiver in the OGS both have cascaded two-stage APT systems.
In the transmitter, the first stage is the satellite attitude control system composed of several flywheels and a capture camera, and the second stage is the fine-control loop formed of a FSM driven by piezo ceramics and a fine camera.
In the OGS, the first stage is the coarse-control loop including a two-axis gimbal telescope and a coarse camera, and the second stage is a similar fine-control loop composed of a FSM and a fine camera.
Detailed parameters and performance of the APT systems are shown in Extended Data Table \ref{tab_s1}.

\begin{figure}[htbp]
\centering
\includegraphics[width =0.9 \linewidth]{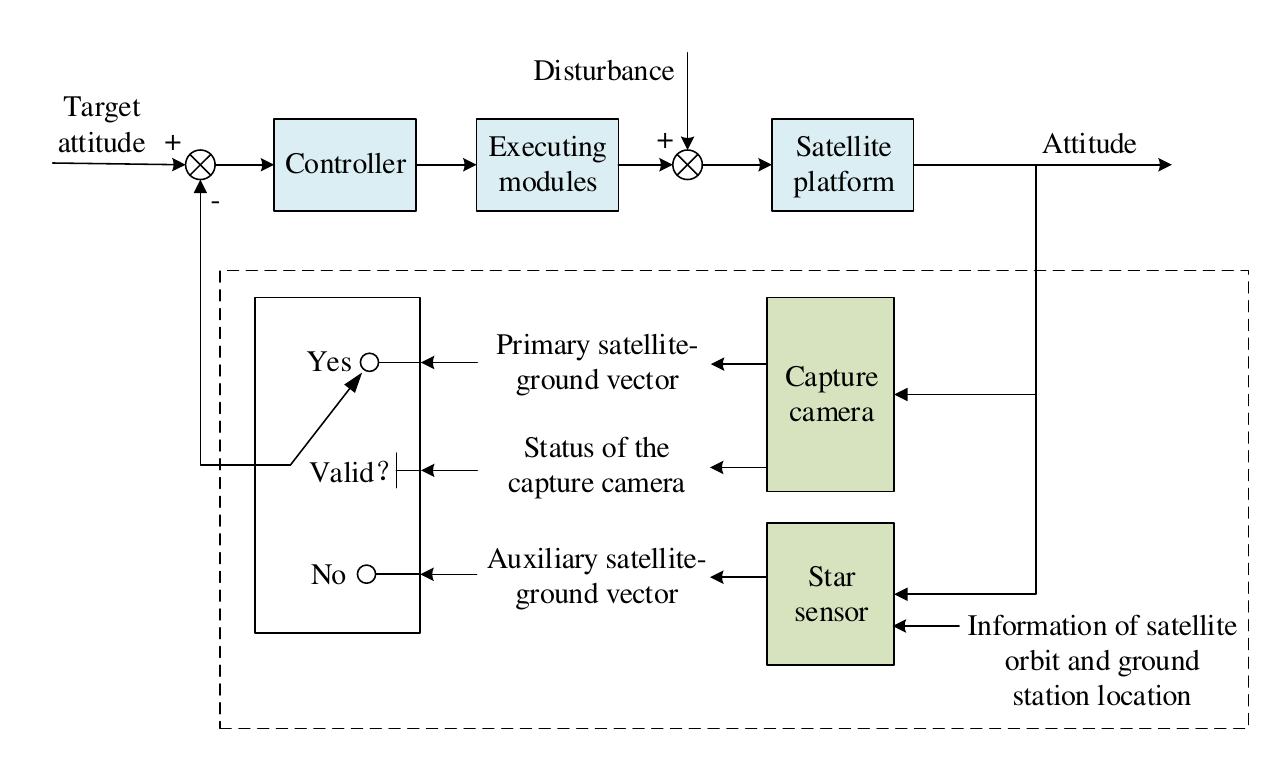}
\caption{
\textbf{Method of satellite attitude control}. 
}
\label{fig_atti_ctrl}
\end{figure}

The method of satellite attitude control is shown in Extended Data Fig. \ref{fig_atti_ctrl}.
To achieve alignment between the capture camera's optical axis and the ground optical station, a unique approach is employed in this satellite's attitude control system. 
Instead of relying solely on traditional three-axis attitude measurement information, this system utilizes the satellite-ground vector information as the control input. 
This vector information is sourced from two different channels: the output from the capture camera, which serves as the primary control input, and an auxiliary satellite-ground vector.
The capture camera's satellite-ground vector information offers exceptional precision ($\mu rad$) and frequency (40 Hz), facilitating high-dynamic and precise attitude control of the satellite. 
The auxiliary satellite-ground vector, on the other hand, serves as a backup control input. 
It operates independently of weather conditions like clouds and rain and ensures that, if the capture camera's satellite-ground vector becomes ineffective due to environmental factors, the satellite can seamlessly transition to using the auxiliary satellite-ground vector to maintain alignment. 
This approach guarantees that the capture camera can swiftly regain valid information and continue autonomously output the satellite's attitude when adverse conditions clear.
The satellite then employs satellite-ground vector information for control and adjusts its attitude using the installed flywheels. 
This closed-loop attitude control cycle operates at approximately 50 ms intervals, corresponding to a control frequency of 20 Hz. 
Finally, through continuous feedback control, the satellite can achieve both high-precision attitude control and coarse tracking towards the OGS.

\subsection{Laser-communication-based key distillation.}
The data post-processing and key-distillation procedure can be distributed into six steps: 
\begin{enumerate}[(1)]
\item True random numbers (RNDs) are generated in real-time using a physical random number generator to prepare the quantum photons.
The prepared quantum photons, along with the downlink communication laser, are transmitted to the OGS.
Simultaneously, the random numbers are also cached for later post-processing at the satellite.
\item Upon receiving the quantum photons and extracting the synchronization patterns, the OGS detects and measures the arrival time of the quantum photons and synchronization patterns.
Ten percent of the quantum photons are sampled to evaluate the QBER, while the remaining ninety percent of the quantum photons are used for the following key distillation.
These measurement results are also cached for later post-processing at the OGS.
The time position information (RND sequence number), the basis information and the raw key of the sampled photons, as well as the time position information (RND sequence number) and the basis information of the remaining photons, are packaged into the original key information and sent to the satellite through the uplink communication laser.
Due to the instability of the satellite-ground classic communication link, the transmitted data may be lost. 
To improve the stability of data transmission and increase the success rate, the data of the original key information will be sent repeatedly frame by frame until the next frame of data is generated.
\item The satellite receives the original key information, reads the previously cached random numbers, and generates the sifted keys and the basis-vector comparison results (basis information and the intensity state information) based on the basis information.
Based on the sampled QBER, the satellite then chooses an appropriate low-density parity check (LDPC) matrix and performs LDPC encoding every 100 kbits of sifted keys packet.
The reason for utilizing the LDPC error-correction method is that it requires the least number of data exchanges between the satellite and the OGS.
Simultaneously, the satellite sends the basis-vector comparison results, along with the generated error-correction syndrome, to the OGS through the downlink communication laser.
\item Upon receiving these results, the OGS reads the previously cached measurement results, and compares them with the basis-vector comparison results to generate the sifted keys and extract the counts of various quantum states.
It performs LDPC decoding according to the received error-correction syndrome, and then calculates the final key length based on the accurate QBER after LDPC error correction and the obtained counts of various quantum states.
Depending on the QKD system's parameters and the satellite's transit orbit, different sizes (100 kbits to 900 kbits with an interval of 100 kbits) for the privacy amplification procedure can be configured to generate the final secure keys.
The OGS then generates the final key generation instruction, signs the uplink data for privacy amplification, and sends them to the satellite.
Due to the instability of the satellite-ground classic communication link, these data are also sent repeatedly until feedback from the satellite is received.
\item The satellite receives the uplink data and the corresponding signature result, and then authenticates the data through the signature.
If the authentication fails, the final key generation will be abandoned.
If the authentication is successful, the satellite completes the privacy amplification procedure based on the final key generation instruction, and extracts the final secure keys. 
Meanwhile, the satellite generates the feedback information and the cyclic redundancy check (CRC) checksum, signs the downlink data for privacy amplification, and sends them to the OGS.
Similarly, these data are sent repeatedly until the next final key generation feedback is generated.
\item 
The OGS receives the downlink data and the corresponding signature result, and then authenticates the data through the signature.
If the authentication fails, the final key generation will be abandoned.
If the authentication is successful, the OGS proceeds with privacy amplification to extract the final secure keys, which are validated using a cyclic redundancy check checksum to verify their consistency.
Finally, the satellite and the OGS share the same secure keys.
\end{enumerate}

\subsection{The microsatellite and the payloads.}
Among the payloads of the microsatellite, the QKD terminal, comprising the integrated QKD light source, weighs around 8.3 kg, while the APT system, which includes the APT control electronics and the APT optical box, accounts for approximately 14.4 kg.
The APT optical box is mounted outside the satellite payload compartment, while the APT control electronics and the QKD terminal are mounted inside the payload compartment.
To achieve temperature stability and ensure good optical performance, the APT optical box is wrapped with thermal insulation materials, combined with active thermal control.
The QKD terminal consists of two physical random number generator chips, a QKD light source, drive electronics of the light source, and an FPGA-based control and key distillation module.
The two optical fiber interfaces of the APT optical box and the QKD terminal are connected by flanges.

Based on the payloads, we adopt a modular approach to realize the satellite.
The platform primarily consists of solar panels, a satellite attitude control module, a microwave antenna, a satellite electronics compartment, a payload compartment, an APT optical box, and several other carried payloads with an overall weight of approximately 12 kg. 
The solar panels and the batteries are used to provide a stable power supply to the satellite and ensure its normal operation.
The attitude control module includes attitude sensors (star sensors), actuators (flywheels), control algorithms, etc., which are used to maintain the proper position and orientation of the satellite in orbit.
The microwave antenna and the transceivers of the satellite are used to facilitate microwave communication with ground satellite monitoring and control stations.
The electronics compartment includes data processors, memory, etc., which are used to process, store, and transmit the satellite data.

\subsection{Comparison with previous missions.}
Comparison with previous missions see Extended Data Table \ref{tab2_comp}.

\subsection{QKD Protocol.}
The final secure keys were extracted using the decoy-state QKD analysis method\cite{Wang2005PRL, Lo2005PRL}.
In the three-intensity BB84 protocol, Alice has two phase-randomized weak coherent state sources in the $X$ basis (namely, the decoy source $x_1$ and the signal source $y_1$), and two phase-randomized weak coherent state sources in the $Z$ basis (namely, the decoy source $x_2$ and the signal source $y_2$), and the vacuum source $o$. 
Each pulse sent by Alice is randomly chosen from one of the five sources (i.e., $l_A = o,x_1,y_1,x_2,y_2$) with probability $p_{l_A}$. 
In our analysis, the intensities were set to $\nu$ for the decoy sources $x_1$ and $x_2$, and $\mu$ for the signal sources $y_1$ and $y_2$, and $w$ for the vacuum sources $o$. 
The probabilities were set to $p_{x_1}=p_{x_2}={p_\nu}/2$, $p_{y_1}=p_{y_2}={p_\mu}/2$ and $p_o=1-p_\nu-p_\mu$. 
In the photon-number space, the states of the pulses from the vacuum decoy and signal sources are respectively given by
\begin{equation}
\rho_{w}=\sum_k a_k \ket{k}\bra{k},\quad \rho_{\nu}=\sum_k b_k \ket{k}\bra{k},\quad \rho_{\mu}=\sum_k c_k \ket{k}\bra{k},
\end{equation}
where
\begin{equation}
a_k=\frac{o^k e^{-o}}{k!},b_k=\frac{\nu^k e^{-\nu}}{k!},\quad c_k=\frac{\mu^k e^{-\mu}}{k!}.
\end{equation}

We denote the counting rate and the error counting rate of source $l_A$ as $S_{l_A}$ and $T_{l_A}$, respectively, and their corresponding error rate as $E_{l_A}=T_{l_A}/S_{l_A}$ for $l_A = o,x_1,y_1,x_2,y_2$. 
The expected values of the counting rate and the error counting rate of source $l_A$ are denoted as $\mean{S_{l_A}}$ and $\mean{T_{l_A}}$, respectively. 
As the number of pulses is finite, we imposed Chernoff bounds on the observed values to estimate the lower and upper bounds.
The Chernoff bound can help us estimate the expected value from their observed values\cite{chernoff1952measure, jiang2017measurement}.

Let $X_1,X_2,\dots,X_n$ be $n$ independent random samples, detected with the value 1 or 0, and let $X$ denote their sum satisfying $X=\sum_{i=1}^nX_i$. $E$ is the expected value of $X$. 
We have
\begin{align}
\label{EL}E^L(X)=&\frac{X}{1+\delta_1(X)},\\
\label{EU}E^U(X)=&\frac{X}{1-\delta_2(X)},
\end{align}
where we can obtain the values of $\delta_1(X)$ and $\delta_2(X)$ by solving the following equations
\begin{align}
\label{delta1}\left(\frac{e^{\delta_1}}{(1+\delta_1)^{1+\delta_1}}\right)^{\frac{X}{1+\delta_1}}&=\xi,\\
\label{delta2}\left(\frac{e^{-\delta_2}}{(1-\delta_2)^{1-\delta_2}}\right)^{\frac{X}{1-\delta_2}}&=\xi,
\end{align}
where $\xi$ is the failure probability. 
Given $X$ and $\xi$, we can get the values of $\delta_1(X)$ and $\delta_2(X)$ by numerical method according to Equation \ref{delta1} and Equation \ref{delta2}.

The upper and lower bounds of $\mean{S_{l_A}}$ and $\mean{T_{l_A}}$ are denoted as $\mean{\overline{S}_{l_A}},\mean{\underline{S}_{l_A}}$ and $\mean{\overline{T}_{l_A}},\mean{\underline{T}_{l_A}}$, respectively, for $l_A =o,\nu,\mu$.
The counting rate of the single-photon pulses was lower-bounded by

\begin{footnotesize}
\begin{equation}
	\underline{s}_1=\frac{p_{\mu}c_1({p_o}^2p_{\mu}(a_0c_2-c_0a_2)a_0\mean{\underline{S}_{\nu}}-{p_o}^2p_{\nu}(a_0b_2-b_0a_2)a_0\mean{\overline{S}_{\mu}}-p_op_{\nu}p_{\mu}a_0(c_2b_0-b_2c_0)\mean{\overline{S}_{0}})}{({p_o}^2p_{\nu}p_\mu((a_0c_2-c_0a_2)(a_0b_1-b_0a_1)-(a_0b_2-b_0a_2)(a_0c_1-c_0a_1)))}
\end{equation}
\end{footnotesize}

and the bit-flip error rate of the single-photon pulses was upper-bounded by

\begin{equation}
	\overline{e}_1=(\mean{\overline{T}_{\nu}}-\frac{p_{\nu}b_0(p_{\nu}b_1\mean{\underline{S}_{0}}-p_0a_1\mean{\overline{S}_{\nu}})}{2p_0p_{\nu}(a_0b_1-a_1b_0)})\frac{p_{\mu}c_1}{p_{\nu}b_1\underline{s}_1}.
\end{equation}

Equivalent lower and upper bounds were calculated and denoted as $\underline{s}_1$ and $\overline{e}_1$, respectively.

Finally, the secure key length $R$ was computed as,

\begin{eqnarray}
R=\underline{s}_1(1-H(\overline{e}_1))-Lec,
\end{eqnarray}

where $Lec$ is information leaked during error correction and $H(x)=-x\log_2x-(1-x)\log_2(1-x)$ is the binary Shannon entropy function.

\subsection{Procedure of satellite-based key relay and encrypted communication.}
\begin{figure*}[htbp]
\centering
\includegraphics[width =1 \linewidth]{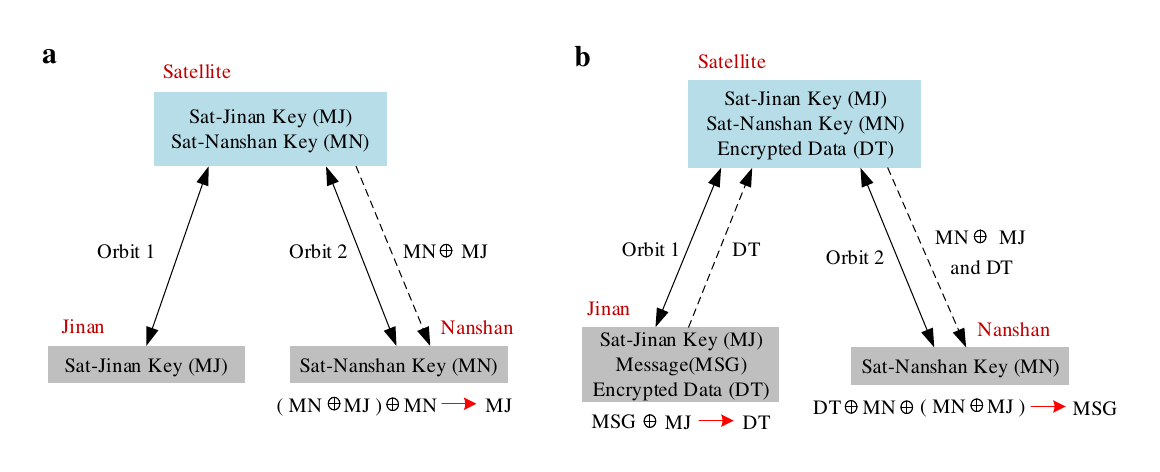}
\caption{ 
\textbf{Illustration of the processes of key relay and encrypted communication between Jinan station and Nanshan station within two satellite orbits.}
\textbf{a}, Key relay.
\textbf{b}, Encrypted communication.
By leveraging the satellite as a trusted relay, both key relay and encrypted communication can be realized between the two OGSs.
}
\label{fig_relay}
\end{figure*}

The key relay process (see Extended Data Fig. \ref{fig_relay}a) can be distributed into three steps: 1, during the first orbit, the satellite completes QKD and shares a secure key (referred to as $MJ$) with Jinan station; 2, during the second orbit, the satellite performs QKD again and shares another secure key (referred to as $MN$) with Nanshan station; 3, once both keys ($MJ$ and $MN$) have been distributed, the satellite performs bitwise exclusive OR operations between them and sends the results ($MN \oplus MJ$) to Nanshan station immediately. 
Nanshan station can employ the key $MN$ to conduct bitwise exclusive OR operation on those keys ($MN \oplus MJ$), resulting in the acquisition of the same shared key $MJ$ with Jinan station.
Finally, through the experiments of two orbits, the key relay between Jinan Station and Nanshan Station was successfully accomplished.

The process of encrypted communication (see Extended Data Fig. \ref{fig_relay}b) follows a similar procedure to the key relay.
The difference lies in the usage of the shared key.
After the satellite and Jinan station complete QKD, they directly utilize the shared key $MJ$ to encrypt the message $MSG$ and transmit the data (referred to as $DT$) from Jinan station to the satellite. 
Then, after the satellite and Nanshan Station complete QKD in the subsequent orbit, the bitwise exclusive OR results between $MN$ and $MJ$, together with the data $DT$, are sent to Nanshan Station.
Nanshan station employs the key $MN$ to conduct bitwise exclusive OR operation on those keys ($MN \oplus MJ$), resulting in the acquisition of the same shared key $MJ$ with Jinan station.
After that, Nanshan station utilizes the key $MJ$ to conduct another bitwise exclusive OR operation on the data $DT$, resulting in the acquisition of the same message $MSG$ with Jinan station.
Finally, through the experiments during two orbits, the encrypted communication between Jinan Station and Nanshan Station was also successfully achieved.

\subsection{Experimental results of satellite-based key relay and encrypted communication.}
The experimental results of satellite-based key relay and encrypted communication are shown in Extended Data Table \ref{tab3_relay}.

\newpage
\clearpage

\clearpage
\clearpage

\clearpage
\begin{table}
\footnotesize
\centering
\caption{
\textbf{Parameters and performance of the acquisition, pointing and tracking (APT) systems}.
At the satellite, the primary beacon light is the 812 nm downlink laser, while the spare beacon light is the 532 nm beacon laser, which has a larger divergence and is mainly used for APT tests at the early stage of satellite launch in orbit. CMOS, complementary metal oxide semiconductor; PZT, piezoelectric ceramic transducer; RMS, root of mean square.
}
\label{tab_s1}
\begin{tabular}{ccc}
\hline
Components & Satellite & Ground station \\
\hline
Coarse pointing mechanism type& Satellite attitude control & Two-axis gimbal mount \\
\hline
Coarse camera type& CMOS & CMOS \\
\hline
Coarse camera field of view& 1.17$^{\circ}$ $\times$ 1.17$^{\circ}$ & 1.69$^{\circ}$ $\times$ 1.27$^{\circ}$ \\
\hline
Coarse camera pixels \& frame rates& 640 $\times$ 640 pixels \& 40 Hz & 2592 $\times$ 1944 pixels \& 14 Hz \\
\hline
Satellite attitude control error &  Range 280$\sim$350 $\mu rad$ & $\setminus$ \\
\hline
Fine pointing mechanism type& PZT fast steering mirror & PZT fast steering mirror \\
\hline
Fine camera type& CMOS & CMOS \\
\hline
Fine camera field of view& \tabincell{c}{512 $\mu rad$ \& 128 pixels \\ 1024 $\mu rad$ \& 256 pixels} & 837 $\mu rad$ \\
\hline
Fine camera pixels \& frame rates& \tabincell{c}{128 $\times$ 128 pixels \& 2000 Hz \\ 256 $\times$ 256 pixels \& 1000 Hz} & \tabincell{c}{256 $\times$ 256 pixels \& \\ 293 Hz} \\
\hline
Fine tracking error (RMS) & 0.55$\sim$1.6~$\mu rad$ & 3.3$\sim$4.0~$\mu rad$ \\
\hline
Beacon laser wavelength& 532 nm and 812 nm & 785 nm \\
\hline
Beacon laser power& \tabincell{c}{30 mW \& 532 nm \\ 2.4 mW \& 812 nm} & 1 W \\
\hline
Beacon laser divergence& \tabincell{c}{17 mrad  \& 532 nm \\ 9$\sim$10~${\mu}rad$ \& 812 nm} & 3.2 mrad \\
\hline
\end{tabular}
\end{table}

\clearpage
\clearpage

\clearpage
\begin{figure}[htbp]
\centering
\includegraphics[width =0.9 \linewidth]{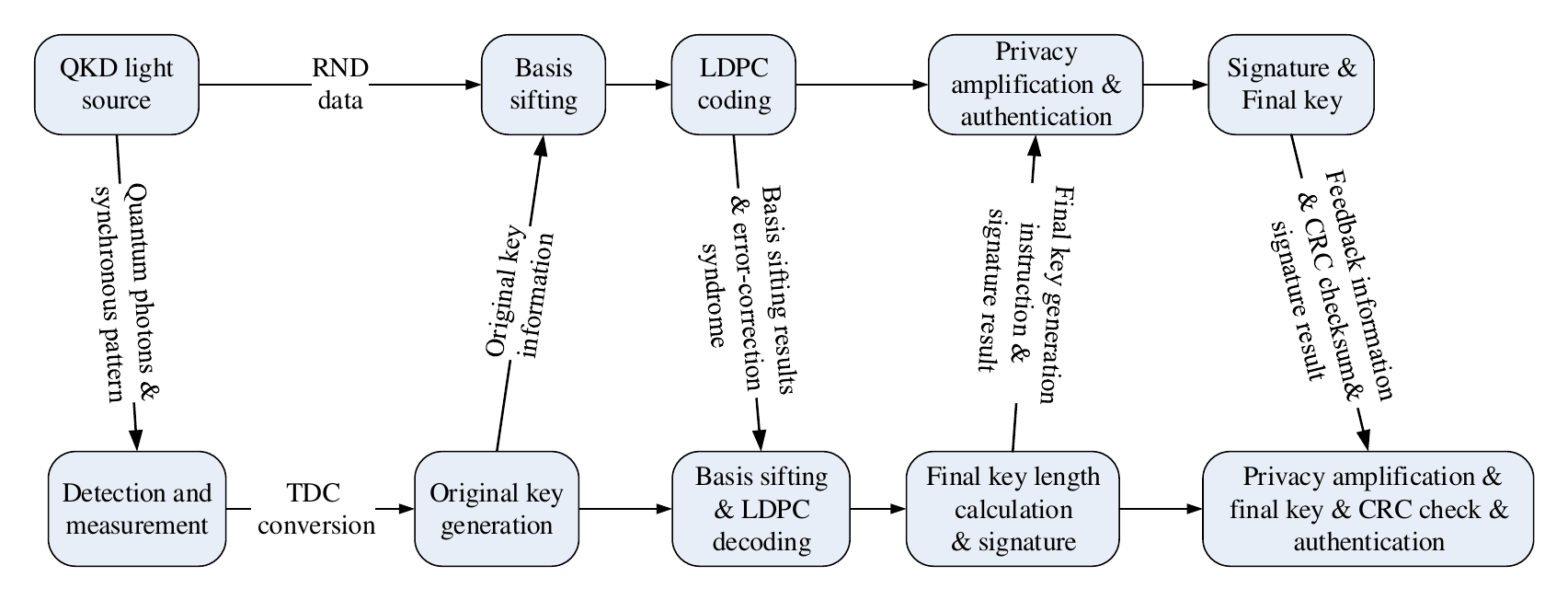}
\caption{
\textbf{Procedure of laser-communication-based time synchronization and key distillation}. 
RND, random number; TDC, time-to-digital converter; LDPC, low-density parity check; CRC, cyclic redundancy check.
}
\label{fig_laser_pro}
\end{figure}
\clearpage

\clearpage
\begin{table*}
\centering
\caption{
\textbf{Comparison with previous missions}.
}
\label{tab2_comp}
\begin{tabular}{ccc}\hline
Parameter & Micius\cite{Liao2017Nature} & This work \\\hline
Payload weight & ${\sim}$250 kg & 22.7 kg \\\hline
Satellite weight & 635 kg & 95.9 kg \\\hline
Transmitter aperture & 300 mm & 200 mm \\\hline
Divergence angle & ${\sim}$10 ${\mu}$rad & 9 $\sim10~{\mu}rad$ \\\hline
Tracking precision (RMS) & $0.6\sim1.5~\mu$rad & $0.55\sim1.6~\mu$rad \\\hline
QKD light source frequency & 100 MHz & 625 MHz \\\hline
\tabincell{c}{Communication scheme \\ for key distillation} &  Microwave & Laser \\\hline
Ground station weight & ${\sim}$13000 kg & ${\sim}$100 kg \\\hline
Key distillation timeliness & 2${\sim}$3 days & Real time \\\hline
\end{tabular}
\end{table*}
\clearpage

\clearpage
\clearpage

\clearpage
\begin{table*}
\centering
\begin{threeparttable}[b]
\caption{
\textbf{Experimental results of satellite-based key relay and encrypted communication}.
$R_{sifted}$ is the packet of sifted keys, $R_{final}$ is the number of final secure keys, and $D_{relay}$ is the number of encrypted communication data between the two OGSs.
}
\label{tab3_relay}
\begin{tabular}{ccccccc}
\hline
Date& Ground station & \tabincell{c}{$R_{sifted}$ \\(*100 kbits)} & $R_{final}$ (bits) & \tabincell{c}{One-time-pad \\ $D_{relay}$ (bits)}& \tabincell{c}{AES-128 \tnote{1} \\ $D_{relay}$ (Byte)}\\
\hline
14 Jul. 2023 &Nanshan & 7 & 96896 &  \multirow{2}{*}{80640}&  \multirow{2}{*}{1M} \\
15 Jul. 2023 &Nanshan & 23 & 382720 &  &  \\
\hline
22 Jul. 2023 &Nanshan & 32 & 519552 &  \multirow{2}{*}{86016}&  \multirow{2}{*}{1M} \\
25 Jul. 2023 &Nanshan & 20 & 447744 &  &  \\
\hline
8 Aug. 2023 &Jinan & 16 & 292992 &  \multirow{2}{*}{87808}&  \multirow{2}{*}{1M} \\
10 Aug. 2023 &Nanshan & 12 & 207616 &  &  \\
\hline
22 Aug. 2023 &Jinan & 20 & 451712 &  \multirow{2}{*}{100352}&  \multirow{2}{*}{1M} \\
24 Aug. 2023 &Nanshan & 20 & 390912 &  &  \\
\hline
1 Sep. 2023 &Wuhan & 33 & 321536 &  \multirow{2}{*}{$\setminus$}& \multirow{2}{*}{1M}\tnote{2}  \\
1 Sep. 2023 &Nanshan & 11 & 114304 &  &  \\
\hline
\end{tabular}
\begin{tablenotes}
\item[1] The generated QKD keys were used as seed keys for the symmetric encryption approach of the Advanced Encryption Standard 128 (AES-128) protocol for encrypting 1 MByte data.
\item[2] Encrypted communication was successfully achieved during two consecutive satellite passes, with a remarkably low latency of approximately 1.5 hours.
\end{tablenotes}
\end{threeparttable}
\end{table*}
\clearpage

\end{document}